\documentclass[11pt]{article}
\usepackage[utf8]{inputenc}

\usepackage{amssymb}
\usepackage{amsfonts}
\usepackage{rotating}
\usepackage{amsmath}
\usepackage{graphicx}
\usepackage{epsfig}
\usepackage{comment} 
\usepackage{graphicx}
\usepackage{cite}

\usepackage[labelfont=bf,labelsep=period,justification=raggedright]{caption}

\makeatletter
\renewcommand{\@biblabel}[1]{\quad#1.}
\makeatother

\date{}

\usepackage[usenames]{color}
\usepackage{url}

\newcommand{\be}{\begin{equation}}
\newcommand{\ee}{\end{equation}}

\newcommand{\bea}{\begin{eqnarray}}
\newcommand{\eea}{\end{eqnarray}}

\begin{document}

% Title must be 150 characters or less
\begin{flushleft}
{\Large
%\textbf{Global Dialect Characterization through Microblogging}
\textbf{Crowdsourcing Dialect Characterization through Twitter}
}
% Insert Author names, affiliations and corresponding author email.
\\
\vspace{0.2cm}
Bruno Gon\c calves $^{1,2,\dagger}$, 
David S\'anchez$^{3}$
\\
\vspace{0.1cm}
\bf{1} Aix-Marseille Universit\'e, CNRS, CPT, UMR 7332, 13288 Marseille, France\\
\bf{2} Universit\'e de Toulon, CNRS, CPT, UMR 7332, 83957 La Garde, France\\
\bf{3} Instituto de F\'{\i}sica Interdisciplinar y Sistemas Complejos IFISC (UIB--CSIC),
E-07122 Palma de Mallorca, Spain\\
$\dagger$ E-mail: Corresponding bgoncalves@gmail.com\\
\end{flushleft}

\begin{abstract} 
We perform a large-scale analysis of language diatopic variation using geotagged microblogging datasets. 
By collecting all Twitter messages written in Spanish over more than two years, we build a corpus from which
a carefully selected list of concepts allows us to characterize Spanish varieties on a global scale.
A cluster analysis proves the existence of well defined macroregions sharing common lexical properties.
Remarkably enough, we find that Spanish language is split into two superdialects, namely,
an urban speech used across major American and Spanish citites and a diverse form that encompasses
rural areas and small towns. The latter can be further clustered into smaller varieties with a stronger
regional character. 
\end{abstract}

%\keywords{Dialectology | Language Variation | Lexical Similarity}

\section{Introduction}

Language is the most characteristic trait of human communication
but takes on many heterogeneous forms. Dialects, in particular, are linguistic varieties which differ
phonologically, gramatically or lexically in geographically separated regions \cite{cha98}.
However, despite its fundamental importance and many recent developments, the way language varies spatially is still poorly understood. 

Traditional methodological approaches in the study of regional dialects are based on interviews and questionnaires administered by a researcher to a small number (typically, a few hundred) of selected speakers known as informants \cite{lab05}. Based on the answers provided, linguistic atlases are generated that are naturally limited in scope and subject to the particular choice of locations and informants and perhaps not completely free of unwanted influences from the dialectologist.
Another approach is the use of mass media corpora which provide a wealth of information on language usage but suffer from the tendency of media and newspapers to use standard norms (the "BBC English" for example) \cite{bau04} that limits their usefulness for the study of informal local variations. 

On the other hand, the recent rise of online social tools has resulted in an unprecedented avalanche of content that is naturally and organically generated by millions or tens of millions of geographically distributed individuals that are likely to speak in vernacular and do not feel constrained to use standard linguistic norms. This, combined with the widespread usage of GPS enabled smartphones to access social media tools provides a unique opportunity to observe how languages are used in everyday life and across vast regions of space.

In this work, we use a large dataset of geolocated Tweets to study local language variations across the world. 
Similar datasets have recently been used to map public opinion and social behavior \cite{borge2011structural,tumasjan2010predicting, culotta2010towards,salathe11-1,salathe12-2,kulshrestha2012geographic,mislove2011understanding,hong_2011}
and to analyze planetary language diversity~\cite{moc13}.% Add GLN reference when available

Preliminary results demonstrating the feasibility of this approach have thus far been limited to considering only few words or just a few geographical areas \cite{eis10,bam12}.
Here, we move beyond the mere proof of concept and provide a detailed global picture of spatial variants for a specific language. For definiteness, we choose Spanish as it is not only one of the most spoken in the world but it has the added advantage of being spatially distributed across several continents\cite{ethnologue,pen00}.
Several other languages such as Mandarin or English have more native speakers or higher supra-regional status but their use is hindered by the limited local availability of Twitter (Mandarin) or a high abundance of homographs that percludes a detailed lexicographic analysis (English).

\section{Methods}

We used the Twitter gardenhose to gather an unbiased sample of all tweets written in Spanish that contained GPS information over the course of over two years. Language detection was performed using the state of the art Chromium Compact Language Detector\cite{chromium} software library.

\begin{figure*}[!ht]
\begin{centering}
\includegraphics[width=1.0\textwidth]{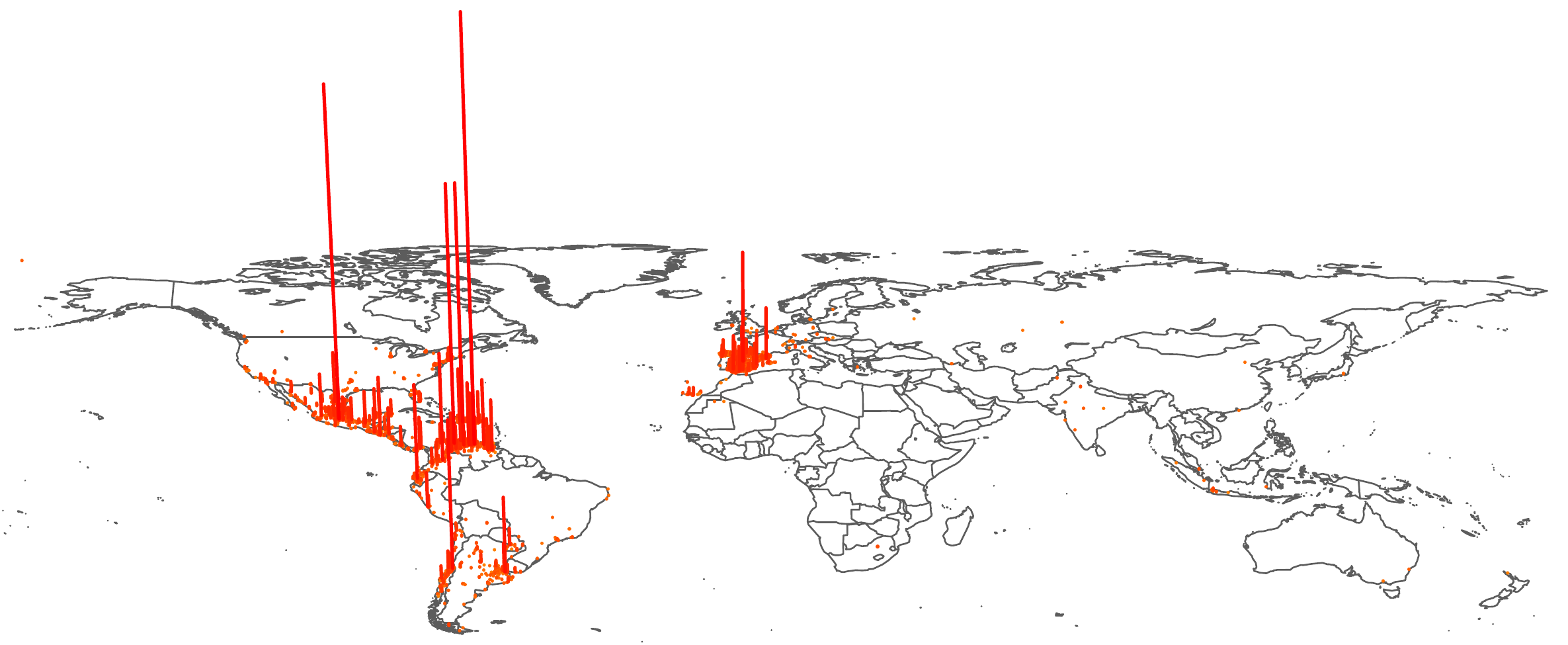}
\caption{\label{fig_map} {\bf Spanish tweet locations.} The overwhelming majority of Spanish tweets are located in Spain and Spanish America but significant contributions arise in certain US states and major Western European and Brazilian cities.}
\end{centering}
\end{figure*}

The resulting dataset contained over $5\times 10^{7}$ geolocated tweets written in Spanish distributed across the world (see Fig. ~\ref{fig_map}). As expected, most tweets are localized in Spain, Spanish America and extensive areas of the United States.
These results are consistent with recent sociolinguistic data~\cite{stewart,moreno}, providing an initial level of validation to our approach. Interestingly, we also find significant contributions from major non-Spanish-speaking cities in Latin America and Western Europe,
likely due to considerable population of temporary settlers and tourists. See Ref. \cite{moc13} for further details and results on this dataset.

Traditional approaches in dialectology have preferred rural, male informants while modern analyses include interactions with urban speakers regardless of age and gender. On average, Twitter users are young, urban\cite{smi10} and more likely to be technologically savvy thus providing more modern perspective on the use of language.

To be able to determine exactly what the major local varieties of Spanish are, we use a list of concepts and utterances selected from an exhaustive study of lexical variants in major Spanish-speaking cities. Reference \cite{varilex} provides a comprehensive list of possible words representing several concepts, such as "popcorn", "car", "bus", etc. We selected a subset of concepts that minimized possible semantic ambiguities by ensuring that they contained no common words\footnote{The complete list of words for each concept studied can be accessed at~\url{http://www.bgoncalves.com/languages/spanish.html}}. 

In our initial set of Tweets we observed $7.5\times 10^{5}$ geolocated instances where words from our catalogue were used. Individual instances were then agregated geographically into cells of $0.25^{\verb+"+}\times 0.25^{\verb+"+}$, which corresponds to an approximate area of $25\times 25$~km$^2$ in the equator.

\begin{figure}[!h]
\begin{centering}
\includegraphics[width=0.45\columnwidth]{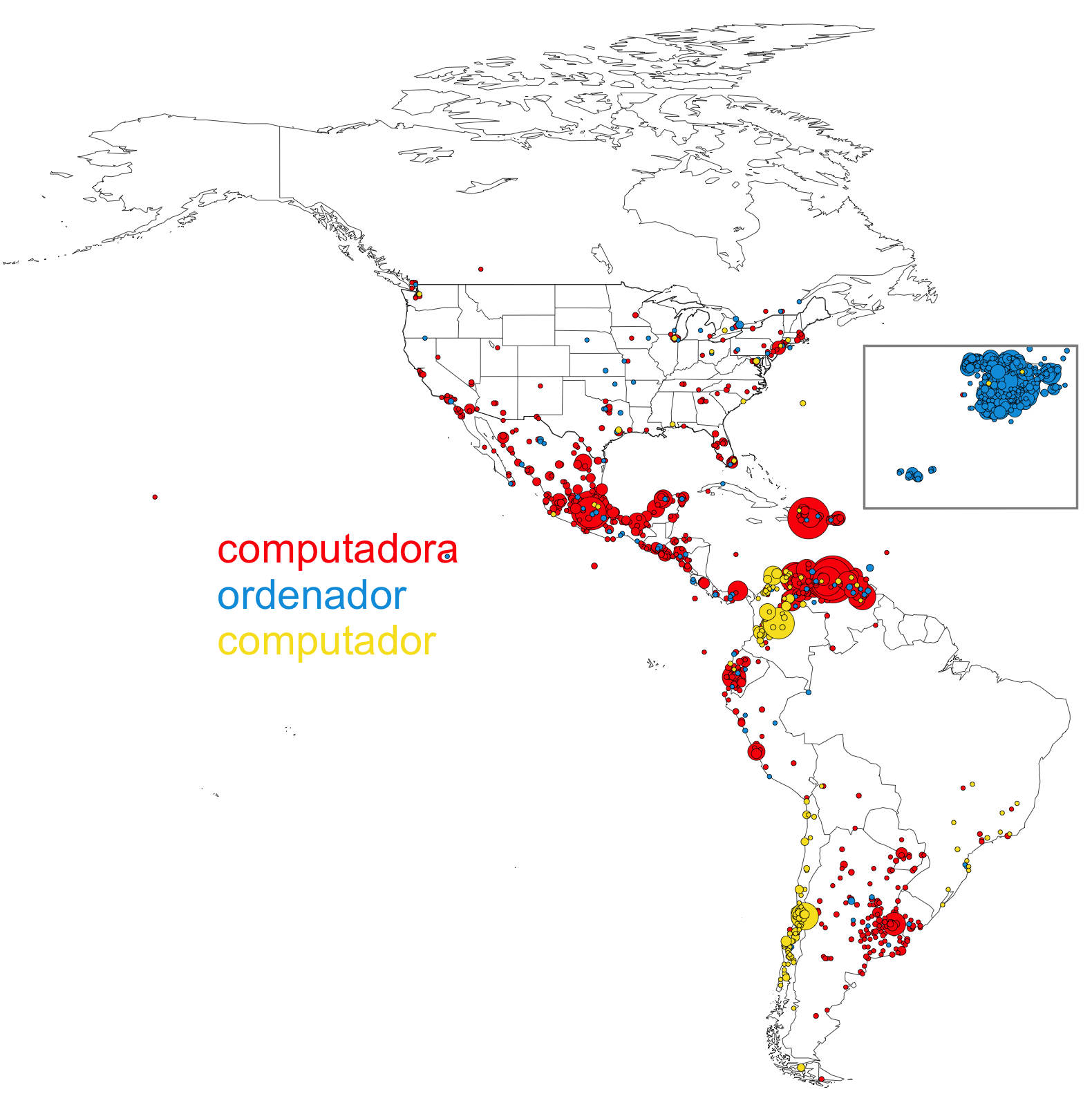}\includegraphics[width=0.45\columnwidth]{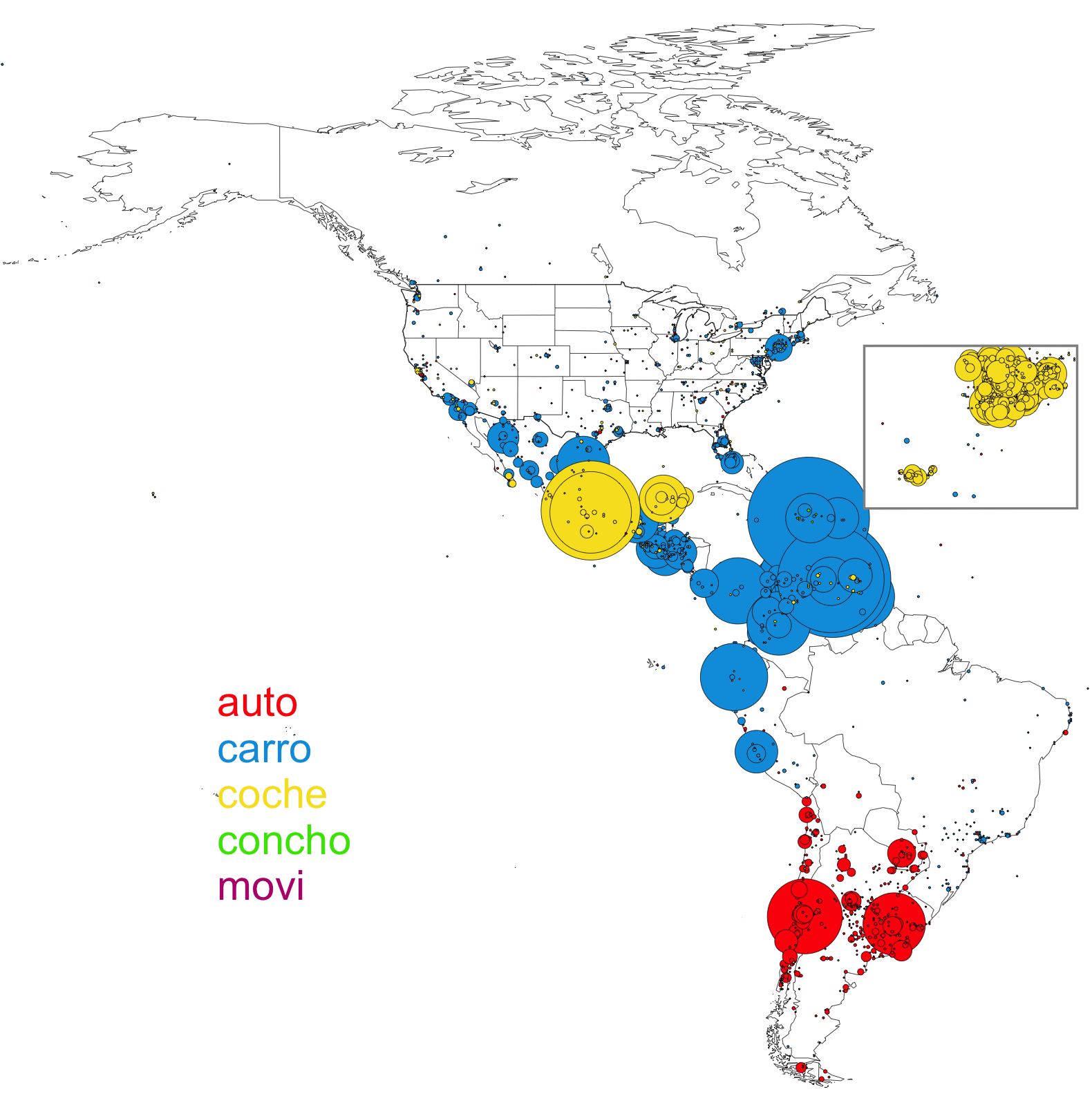}
\caption{\label{fig_popcorn_car} {\bf Geographical distribution of the dominant word for the concepts Computer (left) and Car (right).} Map locations are colored
according to the most common expression found in the corresponding cell. The area of the circle is proportional to the number of tweets}
\end{centering}
\end{figure}

Finally, we define the dominant word for each concept in each geographical cell by a simple majority rule and generate a $M=N_{cells}\times N_{words}$ matrix where element $M_{ij}$ is $1$ when word $j$ is the dominant for a given concept in cell $i$ and $0$ otherwise. The resulting matrix has $N_{cells}=1135$ rows and $N_{words}=131$ columns and constitutes the dataset used for the analysis presented in the remainder of this paper. 

\section{Results and discussion}

Figure~\ref{fig_popcorn_car} illustrates two illustrative concepts ('computer' and 'car') that are both associated to multiple utterances. Each utterance is represented with a different color. We draw a circle centered on each cell with an area proportional to the number of tweets that use the corresponding expression\footnote{The corresponding maps for the other concepts in our catalog can be seen at: \url{http://www.bgoncalves.com/languages/spanish.html}}
It is clear from the map that some expressions (\textit{computadora}, \textit{ordernador}, \textit{computador}) are strongly clustered in space, allowing us to easily define regional dialects characterized by the set of dominant words used to express the concepts in our list. Due to the unique resolution of our data we could limit the isoglosses (boundaries) of the regions corresponding to each concept-word with a high degree of precision. However, the isoglosses corresponding to different concepts can overlap and bundle rendering any simple arrangement of dialect areas almost impossible.

\begin{figure*}[!h]
\begin{centering}
\includegraphics[width=0.45\columnwidth]{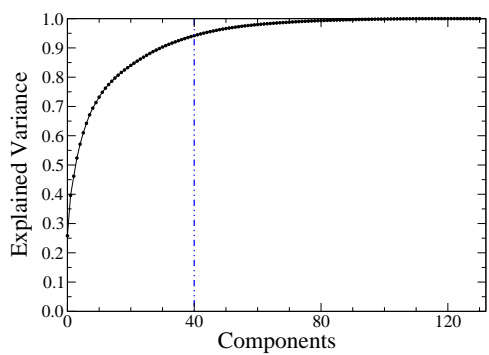}
\caption{\label{PCA} {\bf Cumulative variance explained as a function of the number of components} With $40$ components (vertical blue line) we are able to maintain over $94\%$ of the variance present in the data while significantly reducing the matrix size.}
\end{centering}
\end{figure*}

The natural way to overcome this difficulty and characterize the various regional dialects present in modern day Spanish is to apply machine learning (ML) approaches to automatically cluster the $M$ matrix and identify which cells are closely related to one another. We start by applying Principal Component Analysis to reduce the dimensionality of the matrix $M$. PCA determines the linear combinations of the columns (features in  ML literature) of the matrix that explain most of the variance observed in the rows (observations). We find that by projecting the data onto the $40$ principal components (see Fig.~\ref{PCA}) we are able to maintain over $94\%$ of the variance in the data while reducing by $2/3$ the dimension of the matrix with clear numerical advantages.

\subsection{Superdialects}
\begin{figure*}[!h]
\begin{centering}
\includegraphics[width=0.95\columnwidth]{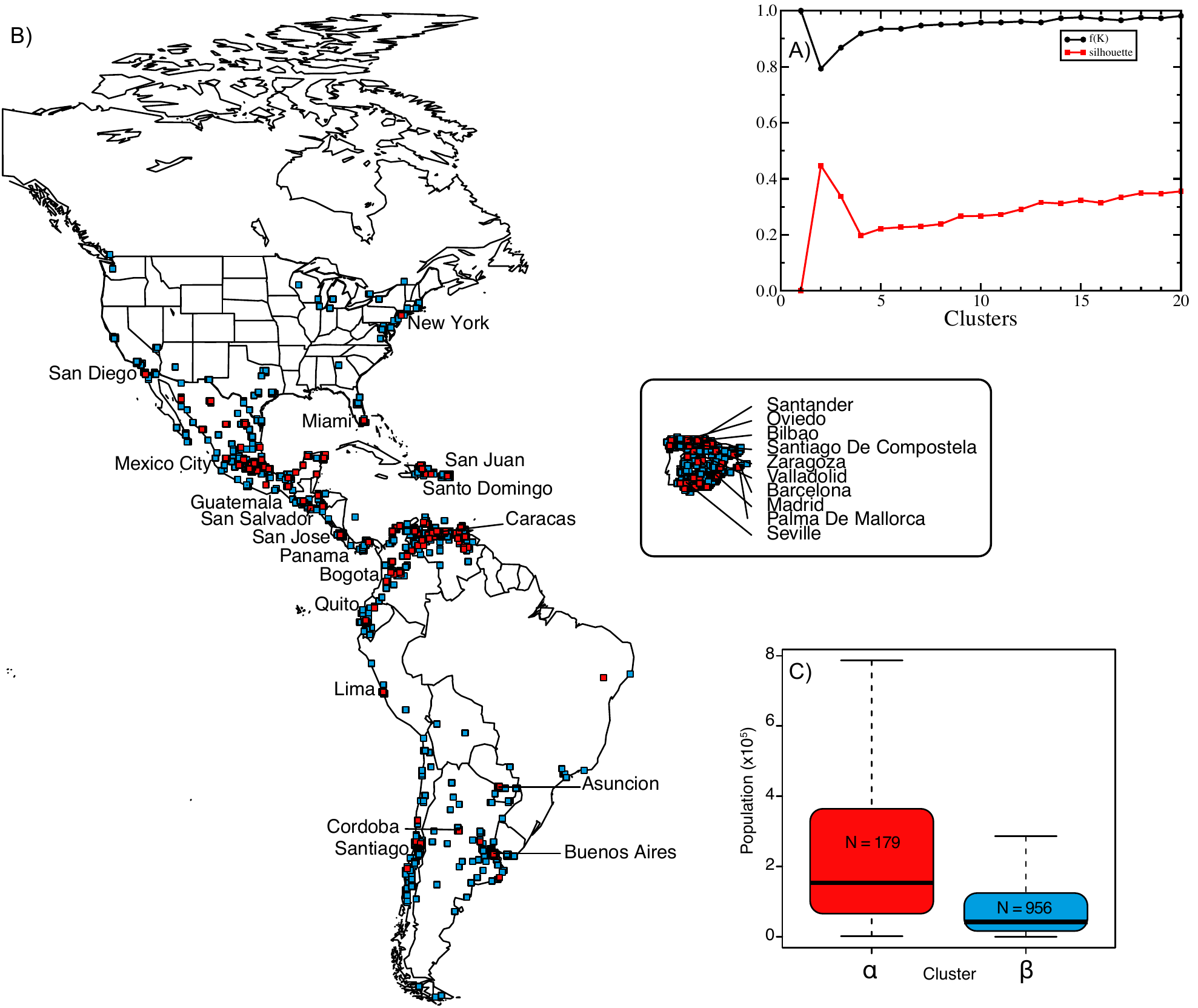}
\caption{\label{population} {\bf Characterization of the two superdialects} A) $f\left(K\right)$ and silhouette statistics as a function of $K$. B) Geographical representation of the two clusters, $\alpha$ (red) and $\beta$ (blue). For visualization purposes we increased the size of each cell. The name of main cities corresponding to superdialect $\alpha$ are shown for clarity. C) Population distribution of the cells corresponding to each cluster.}
\end{centering}
\end{figure*}
The task of identifying meaningful clusters in this matrix is now simplified. We proceed by applying the well known $K$-means\cite{kmeans} algorithm that iteratively refines the position of the centers of $K$ clusters until it finds a stable set of locations. The main dificulty of utilizing this algorithm lies in identifying the correct number $K$ of clusters to utilize. Here, we apply the $f\left(k\right)$ metric introduced by Pham \emph{et al.}  to establish the best value for $K$. We run $K$-means with values of $K$ up to $20$ using $100$ different random initializations and depict the results in Fig.~\ref{population} A). For verification purposes, we also plot the value of the Silhouette~\cite{silhouette} of the clusters found with each value of $K$. Both metrics agree that $2$ is the correct number of clusters (both curves show a extremum at that point), leading to two clusters of size $179$ (cluster $\alpha$) and $956$ (cluster $\beta$), respectively.

A geographic plot of the location of the cells belonging to each  clusters ($\alpha$ and $\beta$) provides a fundamental clue to their meaning (see Fig.~\ref{population} B)).
Strikingly, we find a profound correlation between location of cells belonging to cluster $\alpha$ (red dots) and areas of high population density. We validate this idea using estimates of the population living within each cell provided by the LandScan dataset. Hence, we plot the population distribution boxplot for each cluster in Fig.~\ref{population} C). The results clearly confirm our intuition. Cluster $\alpha$ corresponds to cells with a typical population that is significantly larger than cluster $\beta$. This suggests a natural lexical bipartition of Spanish into two superdialects. Superdialect $\alpha$ is utilized by speakers in main American and Spanish cities and corresponds to an international variety with a strongly urban component while superdialect $\beta$ is comprised mostly of rural areas and small towns.

Our result provides some evidence that the increasing globalization of major languages leads to an homogenization that is especially apparent for the active lexicon \cite{lopez}. Cities (our superdialect $\alpha$)
naturally exert an intrinsic linguistic centripetal force that favors dialect unification, smoothing possible lexical differences. This leveling process present in all countries (thereby its international denomination) is reinforced by the rapid increase of worldwide social ties and the powerful influence of mass media precisely located in important metropolitan areas (Madrid, Mexico City, Miami)~\cite{tru}.Several other sociolinguistic aspects (prestige, higher educational status) also have a role that is more visible in urban environments.

In contrast, rural areas (superdialect $\beta$) are generally more conservative and keep a larger number of characteristic lexical items and native words. As a result, the dialectal area corresponding to superdialect $\beta$ is much more geographically diverse and can be further split, as discussed below.

\begin{figure}[!h]
\begin{centering}
\includegraphics[width=0.85\columnwidth]{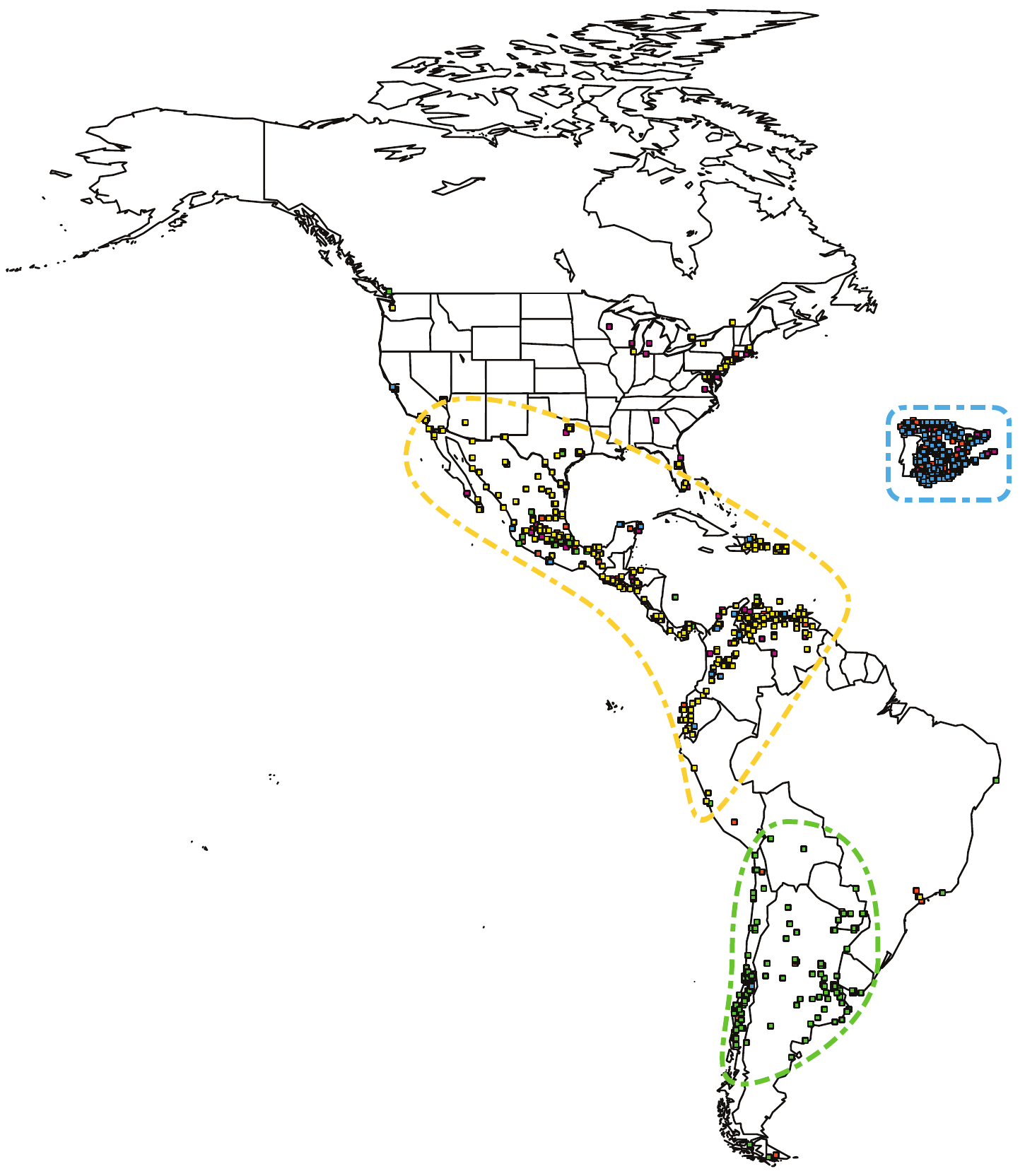}
\caption{\label{cluster} {\bf Characterization of major cluster $\beta$} Geographical representation of regional dialects. For visualization purposes we increased the size of each cell. Three well separated regions are indicated with dashed lines.}
\end{centering}
\end{figure}

\subsection{Regional dialects}

The size imbalance between the two clusters when combined with our intuition suggest that we can also employ the statistical procedure discussed above
to further divide the largest cluster ($\beta$). We apply $K$-means recursively until the remaining cluster has a similar size to the previous ones. In the end, we obtain six well defined clusters that we display in Fig.~\ref{cluster}.
Clearly, three regions can be distinguished. Yellow dots span a wide area covering Mexico, Central America, the Caribbean
and north-western areas of South America. Green dots correspond to the Southern Cone while blue dots are almost exclusively accumulated
within Spain. The first region is quite diverse. In fact, smaller cells can be aggregated into two additional clusters
(depicted with magenta and orange dots in Fig.~\ref{cluster}).
Interestingly, the magenta and orange dots seem to be localized in the Mexican plateau,
the interior of Central America and Andean Colombia, in contrast with the speech of Venezuela,
the Antilles and coastal areas represented with yellow dots. This division
between highland and lowland varieties agrees with classifications
discussed previously in the linguistics literature \cite{cotton}.
 
The two regions marked in Fig.~\ref{cluster} partly reflect the settlement patterns and the formal colonial Spanish
administration within the Empire. Conquerors and settlers occupied first the territories of Mexico, Peru and the Caribbean,
and only much later colonists established permanent residence in the Southern Cone, which stayed away
from prestigious linguistic norms. This strong cultural heritage that can still be observed, centuries later,
in our datasets deserves to be further analysed in future works.

%We can highlight the geographical character of these clusters by assigning each country to the dominant cluster present within its border (while ignoring the international cluster $\alpha$). The resulting map is shown on the right hand side of Fig.~\ref{cluster}. We can easily observe how countries geographically close to each  other teng to gather within the same cluster, thus validating our results. 

%It should also be noted the similarity between this map and that of the "virreinatos" the formal colonial Spanish administrative divisions: The area now belonging to the countries South of Peru were part of the "Rio de Plata" virreinato (with the exception of Chile which had a special status as "Capitania General del Chile". Modern day Ecuador and Peru formed the "Virreinato de Peru" and the area of modern day Venezuela. and Colombia were part of "Nueva Granada" with all the area North of Colombia until Mexico was known as "Nueva Espa\~na". The similarity between these regions and what is found by our empirical approach is striking and points to a strong cultural heritage that can still be observe centuries later.

\section{Conclusions}
Using a large dataset of user generated content in vernacular Spanish, we analyse the diatopic structure of modern day Spanish language at the lexical level. By applying standard machine learning techniques, we find, for the first time, two large Spanish varieties which are related to, respectively, international and local speeches. We can also identify regional dialects and their approximate isoglosses.
Our results are relevant to empirically understand how languages are used in real life across vastly different geographical regions.
We believe that our work has considerable latitude for further applications in the computational study of linguistics, a field full of rewarding opportunities.
One can envisage much deeper analyses pointing the way towards new developments in sociolinguistic studies
(bilingualism, creole varieties). Our work is based on a synchronous approach to language. However, the possibilities presented by the combination of large scale online social networks with easily affordable GPS enabled devices are so remarkable that might permit us to observe, for the first time, how diatopic differences arise and develop in time.

\section{Acknowledgments} We thank I. Fernández-Ordóñez for useful discussions. This product was made utilizing the LandScan 2007™ High Resolution global Population Data Set copyrighted by UT-Battelle, LLC, operator of Oak Ridge National Laboratory under Contract No. DE-AC05-00OR22725 with the United States Department of Energy.  The United States Government has certain rights in this Data Set.  Neither UT-BATTELLE, LLC NOR THE UNITED STATES DEPARTMENT OF ENERGY, NOR ANY OF THEIR EMPLOYEES, MAKES ANY WARRANTY, EXPRESS OR IMPLIED, OR ASSUMES ANY LEGAL LIABILITY OR RESPONSIBILITY FOR THE ACCURACY, COMPLETENESS, OR USEFULNESS OF THE DATA SET.

%\bibliographystyle{plos2009}
%\bibliography{languages}

%
\end{document}